\documentclass{emulateapj}
\usepackage{apjfonts,natbib,epsfig,color}
\usepackage{amsmath,tabularx} 
\usepackage{graphicx}
\newcommand{\beq}{\begin{equation}}
\newcommand{\beqa}{\begin{eqnarray}}
\newcommand{\eeq}{\end{equation}}
\newcommand{\eeqa}{\end{eqnarray}}
\newcommand{\ba}{\[\begin{aligned}}
\newcommand{\ea}{\end{aligned}\]}

\newcommand{\lsim}{\lesssim}
\newcommand{\gsim}{\gtrsim}

\newcommand{\lmk}{\left(}
\newcommand{\rmk}{\right)}

\newcommand{\so}{M_\odot}

\shorttitle{}
\shortauthors{}

\begin{document}
\title{A proposal for enhancing technosignature  search toward the Galactic center}
\author{Naoki Seto}
\affil{Department of Physics, Kyoto University, Kyoto 606-8502, Japan}

\begin{abstract}
We discuss the possibility of  enhancing intelligent life searches toward the Galactic center.  From  the clockwork orbital motions of stars around the Sgr A${}^*$ black hole, we can determine the distance to the Galactic center at an exceptional accuracy, despite  its  remoteness $\sim 8.3$kpc.  In addition, we  can  define precise reference epochs by selecting a prominent object such as the bright B-type star S2.  These properties have  a particular affinity for the coordinated signaling scheme that was hypothesized by Seto (2019) for systematically  connecting intentional senders to searchers without a prior communication. If S2 is actually being used  as a common reference clock,  we can compress the search directions around the Galactic center by more than 2 orders of magnitude,  with the scanning interval of $\sim16$yr. 

\end{abstract}






\section{introduction}

In our Galaxy, an extraterrestrial intelligence (ETI) might  intentionally transmit artificial signals to other unknown  civilizations. However, in spite of our intermittent searches over the past 60 yr \citep{1961PhT....14...40D,2001ARA&A..39..511T,2013ApJ...767...94S,lingam},     we have not succeeded a definite detection  yet. While the number of Galactic  intentional  senders is totally unclear, our observational  and computational resources might be too deficient to  examine the vast parameter combinations that  potentially contain ETI signals \citep{2010SPIE.7819E..02T,2018AJ....156..260W}. 
In  the meantime, there  might be tacit adjustments between the senders and receivers to  compress the parameter space \citep{2018haex.bookE.186W}.  Such adjustments will be beneficial to  both parties for saving various investments,  including the antennas for   signal transmissions and receptions, the electric power for  outgoing signals and data analysis, and so on. 

In the game theory, 
 \cite{schelling} made a pioneering work on  tacit adjustments without prior commutations.  The resultant choice  is  called the Schelling point.  Here,   uniqueness, prominence, and symmetry are considered to be key at  converging the adjustments.
In fact,  some aspects of  ETI search have been  discussed  in relation to  the Schelling point (already mentioned  in \citealt{schelling} about the work by \citealt{1959Natur.184..844C}, {see also} \citealt{1975Natur.254..400P,1977Icar...32..464M,1980Icar...41..178M,2016MNRAS.459.1233K,2018haex.bookE.186W}).
In this context,  the author recently pointed out that a coordinated signaling scheme  (hereafter the concurrent signaling scheme)  might be prevailing in the Galaxy \citep{2019ApJ...875L..10S}.  Through a common usage of a conspicuous astronomical event,
this scheme allows both senders  and receivers to limit the target sky directions,   without depending on their mutual distances (see also \citealt{2018ApJ...862L..21N}).  

In the concurrent scheme, for narrowing down the search directions, we need to   precisely  estimate both  the three-dimensional position and the epoch of the reference event.   The author  proposed to use a future binary neutron star merger for the  scheme \citep{2019ApJ...875L..10S} and past  supernova (SN) explosions for an extended version \citep{2021ApJ...917...96S}.  However, as explained later, these candidates  currently have shortcomings at actual  applications.

Meanwhile, the Galactic center is a salient place in our island Universe (see, e.g. \citealt{1998gaas.book.....B}).  Partly motivated by its specialty,  ETI signals have indeed been searched around the direction of the Galactic center \citep{1985AcAau..12..369S,2017AcAau.139...98W,2021AJ....162...33G,2022PASA...39....8T,2023AJ....165..255S}. Importantly, by observing nearly regular orbital  motions of stars around the massive black hole there, we can determine the distance $l_{\rm H}$ to the Galactic center at an exceptional precision, in spite of its remoteness $\sim 8.3$kpc \citep{2019A&A...625L..10G,2021A&A...647A..59G}.   Furthermore,  we can set  a sequence of reference epochs at high precision, for example,  by using the pericenter passages of a  prominent star  such as the bright B-type star S2 \citep{2019A&A...625L..10G,2021A&A...647A..59G}. These properties  are highly  preferable  for applying the concurrent signaling scheme around the direction of the Galactic center. In this paper, we discuss this possibility,  by concretely setting S2 as the reference clock.

This paper is organized as follows. 
In section 2, we explain the basic idea of the concurrent  signaling scheme  and its extension. We mention drawbacks of  a binary neutron star merger and an SN explosion, which  were  proposed as  conspicuous  reference events in the previous works.   In section 3, we argue the primary aspects  of the Galactic center, in relation to the concurrent signaling scheme. In  section 4,  we evaluate  the expected search directions, using the timing information   of  S2.  In section 5, we discuss the relaxation effects for the long-term regularity of S2's orbit.  In section  6, we discuss issues relevant to our study. Section 7 is devoted to a  short summary.   {In the appendix, we discuss correction effects such as the aberration of light. } 
Throughout this paper, we assume  that the intentional signals propagate at the speed of light $c$.

\newpage
\section{Concurrent Signaling Scheme}

\subsection{basic idea}

We first explain the  basic idea of the concurrent signaling scheme \citep{2019ApJ...875L..10S}. 
As its building block, we can  consider  signal transmissions along a given oriented straight line (indicated in red in Fig. 1). 
Our goal is to synchronize the signal transmissions for all the senders on the line.
To determine the sending  time  at each point  on the line, we use a  conspicuous astronomical event   (e.g.  an  energetic burst) as  a common reference. For each sender, the reference event needs to be observed  in  future and  should be selected on the ground that both its position R and  epoch can be estimate  beforehand  at high precision.   In the next subsection, we discuss a binary neutron star merger as a potential  candidate for a  reference  event.

\begin{figure}[t]
 \begin{center}
  \includegraphics[width=4cm,clip,angle=270]{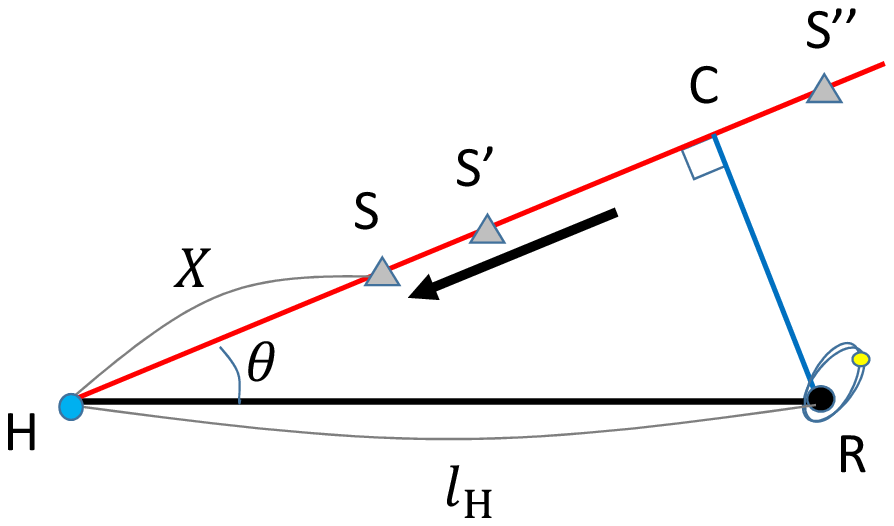}
   \caption{ Schematic picture for the concurrent signaling scheme. The senders S, S' and S'' on the red line can synchronize their signal propagation {so that their transmissions propagate synchronously with an imaginary photon that  {has} passed the closest approach C (to the reference point R) when the event occurs}.  The receiver H can concurrently receive the intentional signals (form the senders S, S' and S'') irrespective of the mutual distances. {We put $X$ as the distance between the receiver H and the sender S.}}
  \label{figure:fig1}
 \end{center}
\end{figure}

 In  three-dimensional  space, for an  arbitrary combination of a  line and an external  reference point R, the closest approach  C will  be the unique point selected as the Schelling point (see Fig. 1). Then, we can synchronize the signal transmission along the oriented  line so that the emitted signals pass the close approach C at the time of the burst occurrence. { We illustrate the time sequence of the signal propagation in Fig. 2.   } Strictly speaking, among the senders S, S' and S'' in Fig.  1,  only the signals of the sender S'' (upstream of the point C) can  pass the point C.  But the remaining senders S and S'  (downstream of C) can easily know the appropriate epochs  for their signal transmissions along the oriented  line (see panel (c) in Fig. 2).    Note that, in the present  scheme, a sender  should  receive upstream signals, at the time of transmitting its own intentional signals   (toward the antipodal direction). 
  On the red line  in Fig .2, at each epoch,   the position of the intentional signal (blue arrow) is identical to the position of the sender.   

Next,  as an example,  we consider an ETI searcher H at the distance $l_{\rm H}$  from the reference.   We discuss  how its receiving directions change with time.  At the time $2l_{\rm H}/c$ before observing reference event, the receiving direction is  { $\theta=\pi$}  in Fig. 1 (considering the limit $\rm lim ~RC\to 0$).   Similarly,  just at observing the  reference  event, the receiving direction is {$\theta=0$}.    In the time between, the receiving directions are  on a circle in the sky, centered by the reference direction.  Its opening angle $\theta$ can be determined  as a function of time   (presented later in  Section 4 with the parameter $q=0$), only from the information of the reference.   
In  this manner,  the concurrent signaling scheme is simple and  enables involved civilizations to largely compactify the sending and receiving  sky directions, irrespective of their mutual  distances.

\subsection{Reference  Events}
As  mentioned in the previous subsection, for  the common reference,  we need to select a  future astronomical event.  Its position and epoch must be predicted beforehand  at high precision. 
As a candidate of such  a reference in our Galaxy,  \cite{2019ApJ...875L..10S} proposed a binary neutron star merger (at a typical distance $l_{\rm H}\sim 10$kpc)  with the observed  orbital  frequency  $\sim 1.5$mHz.  From the first principles of physics, 
by using the planned Laser Interferometer Space Antenna (LISA; \citealt{2017arXiv170200786A})  for $\sim 10~$yr, its distance $l_{\rm H}$ can be determined at subpercent level with relatively negligible error for the arrival  epoch of the merger signal  \citep{2019ApJ...875L..10S}. 
However, LISA is unlikely to be launched before the mid-2030s.  In addition, we might not have a suitable  neutron star binary  in our Galaxy, given the decline of the estimated  comoving merger rate  by the LIGO-Virgo-Kagra  collaboration in the past 4 yr \citep{2021ApJ...913L...7A}.    More specifically, there might be no neutron star binary  whose merger signal will arrive to us in the time $2l_{\rm H}/c$ from now.  It  is thus interesting to think about other  references and/or possible extensions of the scheme.

\begin{figure}[t]
 \begin{center}
  \includegraphics[width=9cm,clip]{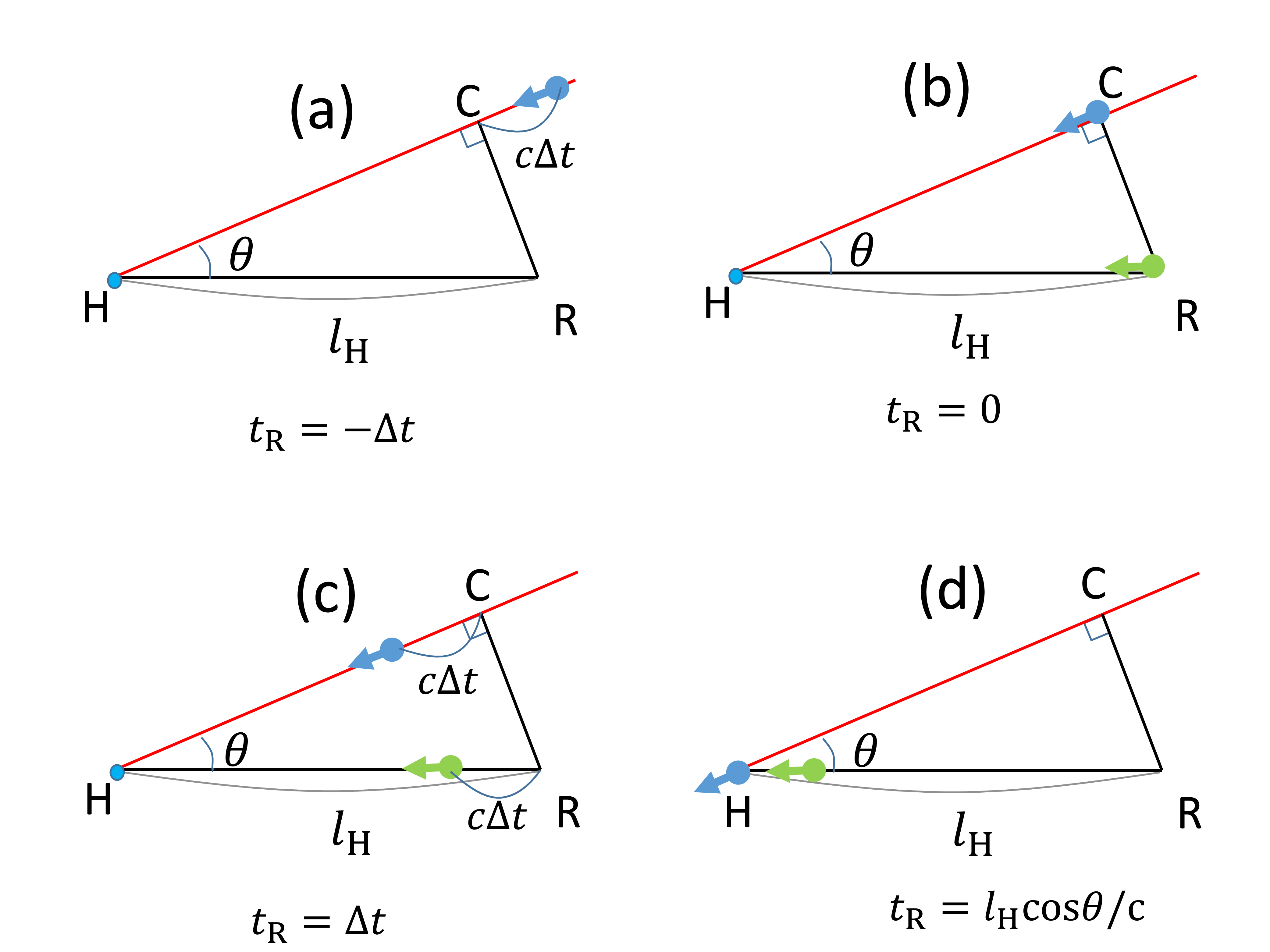}
   \caption{{The time series of the signaling scheme with $q=0$. We put $t_{\rm R}=0$ at the time of the reference event.  The blue arrow shows the propagation position of the intentional signals and is identical to the position of the sender at the time. The green arrow shows the position of the photons marking the  reference event.   At the time $t_{\rm R}<0$, the intentional signal  is at the upstream of the close approach C (panel (a)). The intentional signal reaches C at $t_{\rm R}=0$ (panel (b)).    With CH<RH, the intentional signal is observed earlier than the photons of the  reference event (panel (d)).  }}
  \label{figure:fig1}
 \end{center}
\end{figure}
For  the original scheme in Seto (2019),  we need  to use a reference astronomical event that will be observed in the future.  As an alternative choice  for  the  synchronized passage time (of the intentional  signals) at  the point C,  Seto (2021) proposed  the epoch when the reference event signal reaches the point C (namely, ${\rm RC}/c$ later than the original epoch with $q=0$).     Then, from the triangular inequality in Fig. 1
\begin{align}
\rm RH<RC+CH, 
\end{align}
 we can receive (and send)  intentional signals after observing the reference astronomical event. In Section 4,  this extension will be  analyzed  with the parameter  $q=1$. 
 
  While the additional complexity will not  be  preferred for the tacit adjustment,  this post-factum nature enables us to use historical SN explosions recorded in the past $\sim 2000$ yr,  as  potential references  for the concurrent scheme.      
 Seto (2021) provided the receiving sky directions associated with five SNe whose remnants are relatively well identified.  Using Gaia data, \cite{2023AJ....166...79N} actually examined  potential ETI signals for some of the predicted directions (also adding the case for SN1987).  

 Unfortunately, even though the arrival   times of the SN signals are well determined from the historical records, their distances  typically have large uncertainties (e.g.  20\% for the Crab pulsar associated with  SN1054; see \citealt{2008ApJ...677.1201K}).   Consequently, the angular width $\Delta \theta$ of the receiving circle becomes large and the compactification of the directional parameters is limited.

\section{Galactic Center}

The Galactic  center is a cynosure place in the Milky Way, and the surface density of stars is highly enhanced around its sky position \citep{1998gaas.book.....B}.  
In consideration of these aspects,  ETI signals  have been searched around the direction of the Galactic center \citep{1985AcAau..12..369S,2017AcAau.139...98W,2021AJ....162...33G,2022PASA...39....8T,2023AJ....165..255S}. Notably,  in relation to the concurrent signaling scheme,  the distance to the Galactic center (more preciously Sgr A$^*$) is measured  at an exceptional accuracy despite  its remoteness  \citep{2019A&A...625L..10G,2021A&A...647A..59G}. This measurement is based on the nearly clockwork  orbital motions of stars around the massive black hole there.     For   a  given orbital period (e.g. 10yr),  because of its extraordinary mass ($\sim 4.2\times 10^6\so$),   the Sgr A$^*$ black hole swings nearby stars  at much larger spatial scale, compared with ordinary main-sequence star binaries. Thus, the combination of their radial velocities  and the proper motions allows us to make a high-precision  measurement of  the distance $l_{\rm H}$, on the basis of the simple orbital dynamics.  Indeed,  using the  Very Large Telescope Interferometer,  \cite{2021A&A...647A..59G} recently reported
\begin{align}
l_{\rm H}=8275\pm9_{\rm stat}\pm 33_{\rm sys}~{\rm pc}. \label{eqq}
\end{align}

How about the reference time of the Galactic center for our signaling scheme?  Here the nearly regular orbital motions of the stars will be very useful. For each star, a pericenter passage  time will be the primary candidate for the reference  epoch.   In fact, because of the regularity of an orbit, we can calculate the sequence of its passage  times both in the past and future directions.   Note that, compared with an apocenter passage,  a pericenter passage has a more drastic orbital variation,  in particular for a highly eccentric orbit.

 However, it is  not obvious which star in the cluster we  should choose for  a reference.  Here, from a surveillance of relevant research activities by human beings in the past $\sim 20$\,yr (see, e.g. \citealt{1996Natur.383..415E,2002Natur.419..694S,2003ApJ...586L.127G,2010RvMP...82.3121G,2016ARA&A..54..529B}),  we select the B-type star S2, which is regarded as  the most prominent star for its luminosity and orbital period.  In addition, the formation scenario of this bright star was  actually 
puzzling at its discovery \citep{2003ApJ...586L.127G}.  At the solar system barycenter (SSB), this star has the most recent pericenter passage at 
\beq
t_0=2018.378990\pm 0.000082  ~{\rm yr}
\eeq
 with the estimated orbital  period 
 \beq
 P_0=16.0458\pm 0.00013 ~{\rm yr}
 \eeq  
 and a high eccentricity $e=0.8842$ \citep{2021A&A...647A..59G}.

Our choice of S2 is partly for concrete demonstration of the scheme and might be also  useful  for starting reanalysis on the already obtained SETI data around the Galactic center direction.   However, we should be open-minded to consider the possible references other than S2, in particular intriguing  objects  uncovered in the years to come.

\section{Search Directions}
We now evaluate the opening angle $\theta$ of the receiving direction around the Galactic center ($\alpha={\rm 17h46m40s}, ~\delta=-29^\circ0'28''$), with the time reference determined by  the pericenter passages of S2.
We first explain the parameter $q$ defined for  the choice of the passage time of intentional signals at the closest approach C in Fig.  1. As already mentioned in Sections 2.1 and 2.2,  {the parameter $q$ specifies the epoch  when the intentional signals pass the closest approach C.   For the original choice $q=0$, the epoch is identical to the occurrence of the reference event.  For $q=1$,  the epoch is when the photons of the reference event reach  the point C (i.e. RC$/c$ later than the choice $q=0$).}

In  terms of the timing parameters $t_0$ and $P_0$  of S2,  we can approximately put the epochs of its  future and past pericenter passages (observed at the SSB) by   
\begin{align}
t_{n}& \simeq t_0+P_0 n  \label{eq}
\end{align}
with the integers  
$n\le 0$ for the past ones (related to $q=1$) and  $n\ge 1$ for the future ones (related to $q=0$). 
For a  given $n$, we estimate the opening angle $\theta_n$  of the receiving cones, as functions of the data taking epoch $t_d$.   Below, we collectively deal with different $n$.  If this abstract treatment looks confusing, one can fix the integer $n$ at a specific value (e.g. $n=2$ automatically with $q=0$).

 For each passage  time $t_n$, we  take into account the photon travel times for the distances ER, CE, and RC (only for $q=1$)  shown in Fig. 1. We obtain the condition for $\theta_n$ 
\begin{align}
t_d=t_n-l_{\rm H}/c+l_{\rm H}\cos\theta_n/c+ql_{\rm H}\sin\theta_n/c \label{bes}
\end{align}
or equivalently
\begin{align}
(1-\cos\theta_n-q\sin\theta_n)=fn-f(t_d-t_0)/P_0  \label{base}
\end{align}
Here we define the ratio 
\begin{align}
f\equiv \frac{P_0c}{l_{\rm H}}
\end{align}
with $f=6.02 \times 10^{-4}$ for S2 \citep{2021A&A...647A..59G}.

Next, just for roughly showing  the dependence on  $n$, we put $t_d=t_0$ and solve Eq. (\ref{base}), assuming $\theta_n\ll 1$.  
For the original scheme with $q=0$, we obtain  
\begin{align}
\theta_n\sim \sqrt{2n f} \label{th0}
\end{align}
with $n>0$.  Meanwhile,  for the modified version with $q=1$, we obtain 
\begin{align}
\theta_n\sim |n|f  \label{th1}
\end{align}
with $n\le 0$.

\begin{figure}[t]
 \begin{center}
  \includegraphics[width=7.9cm,clip]{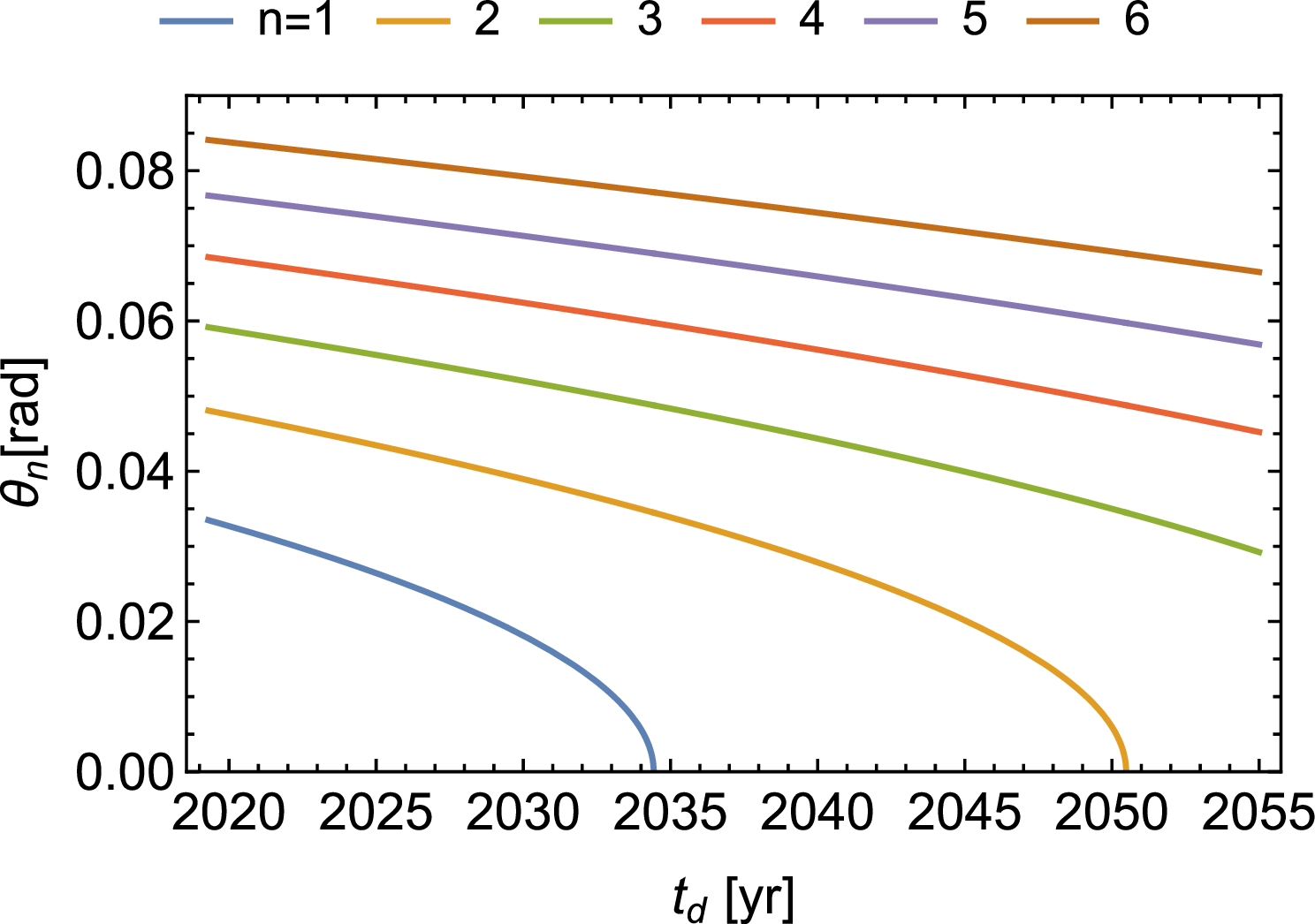}
   \caption{ The opening angles $\theta_n$ for the receiving  cones with $q=0$ around the direction to the Galactic center.  For the  reference time, we use the periastron passage of S2 around Sgr $\rm A^*$  ($t_0=2018.38$ and $P_0=16.046$~yr). Two solutions $\theta_1$ and $\theta_2$  disappear at the future pericenter passages in $t_d=2034$ and 2050. }
  \label{figure:fig2}
 \end{center}
\end{figure}

\begin{figure}[t]
 \begin{center}
  \includegraphics[width=7.9cm,clip]{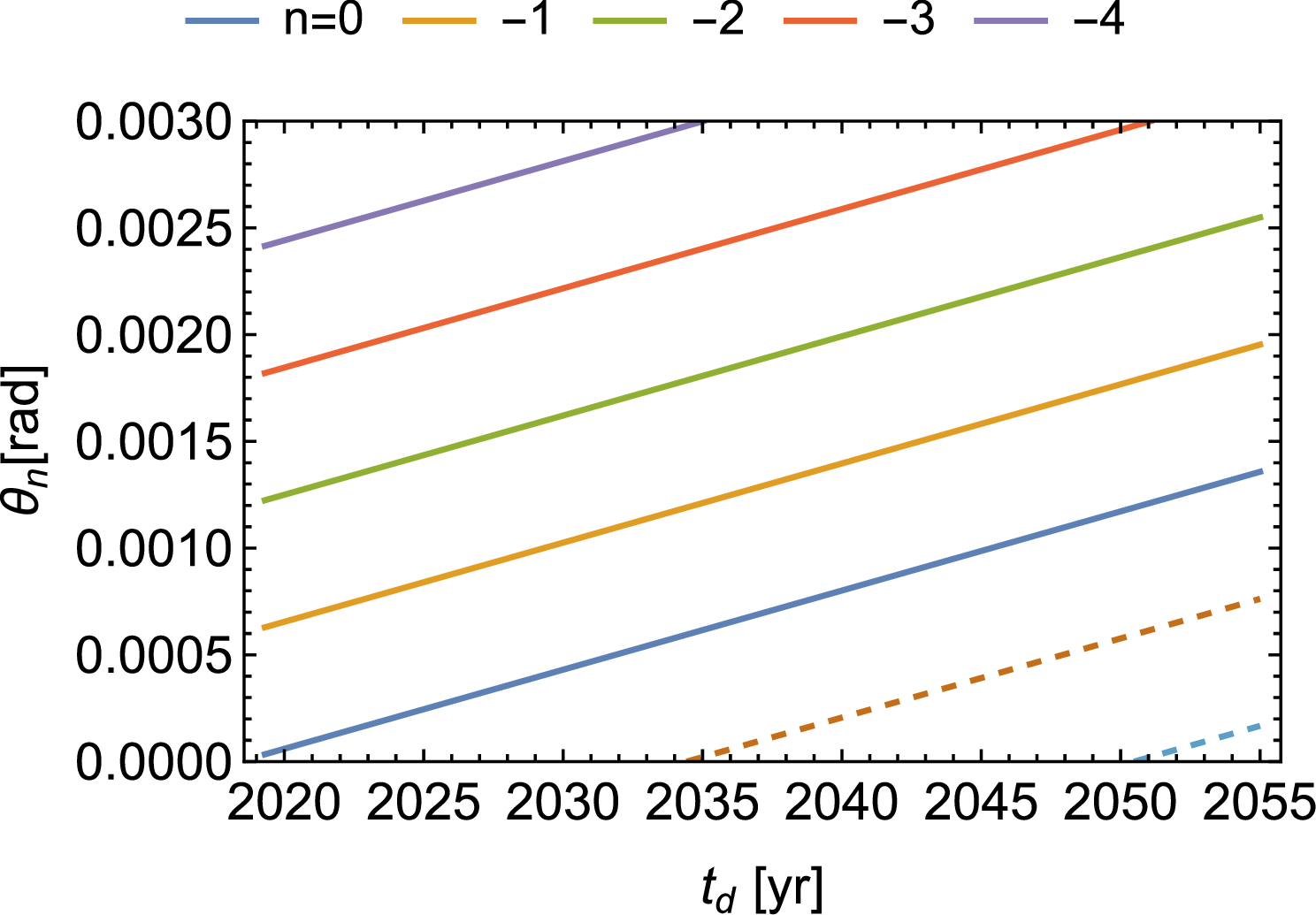}
   \caption{ {The opening angles $\theta_n$ for the receiving  cones with $q=1$ (based on S2).  New solutions appear at the future pericenter passages in $t_d=2034$ and 2050. }  }
  \label{figure:fig2}
 \end{center}
\end{figure}

For a moderate magnitude  $|n|$ (e.g. $\lsim 10$), we can have a much  larger angle $\theta_n$ with the original choice $q=0$ than the extended one $q=1$. Also considering the lucidity of the original method, we mainly discuss the choice $q=0$ ($n>0$) hereafter.   Then, as a function of the data taking epoch $t_d$, the formal solution to Eq.  (\ref{base}) is written  as
\begin{align}
\theta_n=\arccos [1-fn-f(t_d-t_0)/P_0]
\end{align}
with $n>0$. 

In Fig. 3, we present the numerical results for $n=1,\cdots,6$.    We have the zero-points $\theta_n=0$ at $t_d=t_n$ and  the horizontal  periodicity  of $P_0=16.0$\,yr. The shapes of the curves are identical and independent of $P_0$ and $t_n$. Note that, even with $n=6$, we can cover the sky area with $\theta_6=0.08\,{\rm rad}\sim 5^\circ$.  { For a reference, in Fig. 4, we also plot the results  for $q=1$. As shown in Eqs. (\ref{th0})  and (\ref{th1}),  they are much smaller than those in Fig. 3  for $q=0$.  }

It is important  to recall that,  only by  precisely measuring the three parameters $l_{\rm H}$, $t_0$ and $P_0$,   we  can evaluate the opening angle $\theta_n$ for the signal reception and transmission. Then, we can simply  access to the  coordinated signaling scheme without prior communication.  This is the major advantage of the scheme. 

In Eq. (\ref{eq}),  we put $nP_0$ for the next $n$ rotation periods.
If this estimation  has a relative error smaller than that of the distance $\Delta l_{\rm H}/l_{\rm H}$, the width for each receiving circle $\Delta \theta_n$ is given as 
 \begin{align}
\Delta \theta_n\sim \theta_n \frac{\Delta l_{\rm H}}{2l_{\rm H}}, \label{et}
\end{align}
{corresponding to timing error of }
\beq
\Delta t\sim n P_0 \frac{\Delta l_{\rm  H}}{l_{\rm  H}}. \label{ert}
\eeq
For $n=6$, 
we obtain $\Delta  \theta_6\sim 2\times 10^{-4}\,{\rm rad}\sim0.7'$,  using the current measurement accuracy $\Delta l_{\rm H}/l_{\rm H}\sim 0.005$. We can compactify the search directions by more than a factor of 100.

Note that the pericenter distance of S2 is $\sim17$ light-hours.  In Eq. (\ref{base}), this should be  compared with Eq. (\ref{ert})  of $\sim 30n$ light-days.  Therefore, for the accuracy limited by the distance error $\Delta l_{\rm H}/l_{\rm H}$, we will be able to safely ignore the positional difference between  the massive black hole and the pericenter of S2 (even considering the subtle general relativistic effects).

\section{deviation of clock}

The scheme in the previous section relies on the regularity of the orbital motion of reference stars (in particular the periods between the  pericenter passages). 
In  reality, the  orbital elements of stars in the nuclear star clusters will be gradually out  of order, mainly by gravitational interaction with neighboring stars (see, e.g. \citealt{2013degn.book.....M}).  Using the   energy relaxation time  $T_{\rm rel}$ ($\sim 10^{10}$yr for S2; see, e.g. \citealt{2010RvMP...82.3121G}) for the variation of the semimajor axis, 
we   crudely estimate the period fluctuation $\Delta P\sim ( T/T_{\rm rel})^{1/2}P$ in the time interval $ T$.  For $ T\sim 100$yr  (corresponding to $n\sim6$), we have $\Delta P/P\sim 10^{-4}$,  which is smaller than the current distance error $\Delta l_{\rm H}/l_{\rm H}\sim 5\times 10^{-3}$. 
Given the chaotic nature  of the orbital  dynamics and probable existence of dark objects (e.g. stellar mass black holes),  it would be difficult to beforehand correct the relaxation effects at high precision.

\begin{figure}[t]
 \begin{center}
  \includegraphics[width=7.9cm,clip]{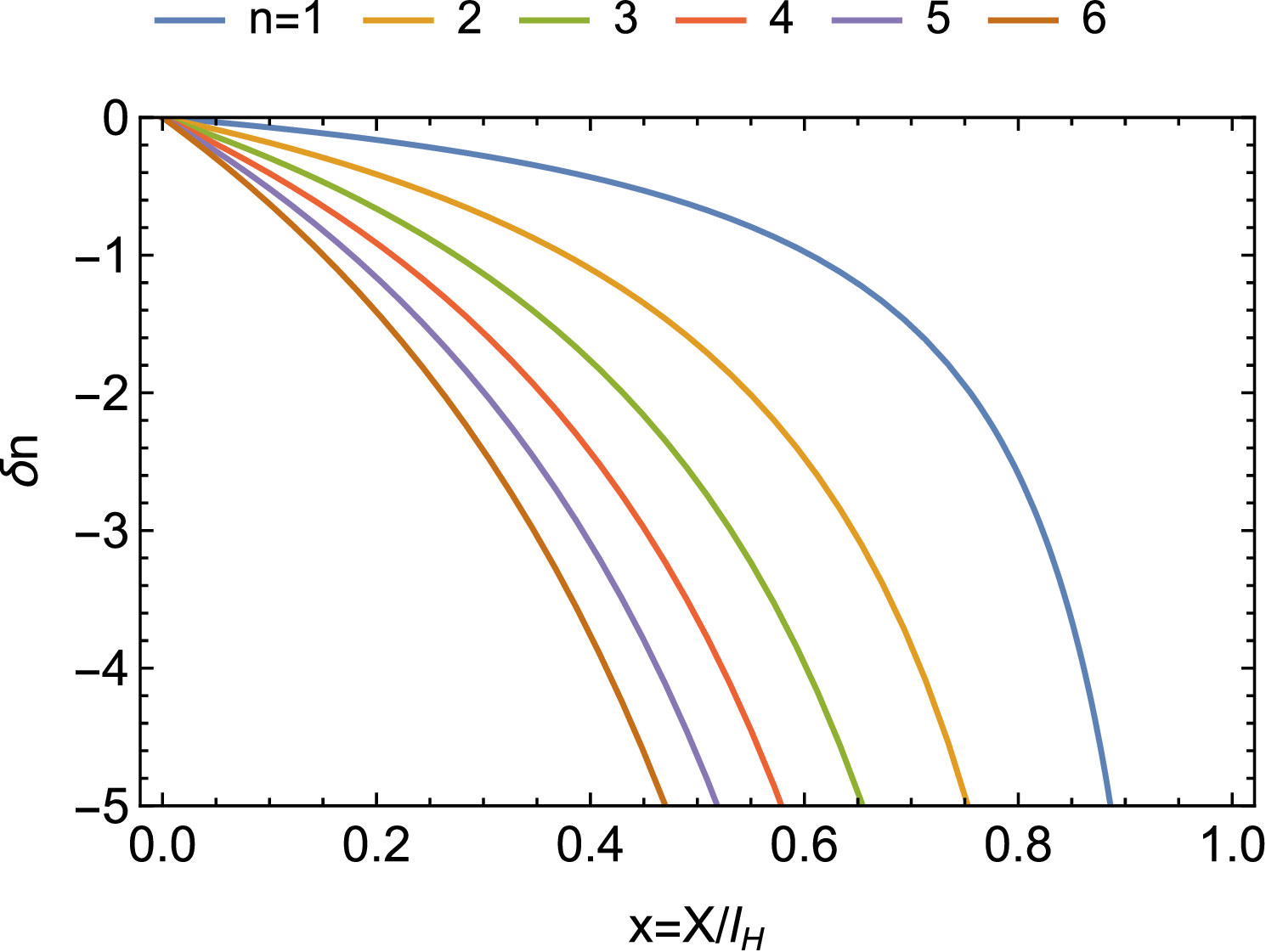}
   \caption{ The rotation cycle difference $\delta n$ of S2 observed by  the Earth and  a sender $S$ at the distance $X$ (see  Fig. 1). We set our observational epoch at $t_d=2024.000$ {and the difference $\delta n=5$ corresponds to the time difference of $\sim 80$\,yr.} 
 }
  \label{figure:fig3}
 \end{center}
\end{figure}

In any case, it would be reasonable for involved civilizations  to calibrate the reference times $t_n$ in Eq. (\ref{bes}) for the signaling, based  on the actual data of the recent pericenter passages.  Here, in Fig. 1, let us consider a signal transmission from the sender  S to us (H) under the concurrent scheme. 
We should notice that, even with respect to  the common S2 clock, the observed orbital phase at the signal transmission (by S) is different from that at the signal reception (by H).  Considering the long-term deviation of the S2 clock, we would like  to  have a small  gap for the intervening orbital cycles.  Here, we  evaluate the cycle gap  $\delta  n$ in a stepwise manner. 
We put the  transmission  time at the sender S by $t_S$ and its distance from   the Galactic center  by $l_S$.   By comparing the time difference back at the Galactic center (namely $t_S-l_S/c$ for  the sender and $t_d-l_{\rm H}/c$ for us), the cycle gap  $\delta n$ is formally given as
\begin{align}
\delta n=\frac{(t_S-l_S/c)-(t_d-l_{\rm H}/c)}{P_0} .\label{dn}
\end{align}
{This expression is valid also for the choice $q=1$.}
Next we apply this formal expression to the  sender S on the propagation line in Fig. 1 at the distance $X$. The sending time $t_S$ of S is given by 
\begin{align}
t_S=t_d-X/c,
\end{align}
and its Galactic radius   becomes
\begin{align}
l_S=\sqrt{l_{\rm H}^2+X^2-2l_{\rm H} X\cos\theta_n}.
\end{align}
Plugging in these expressions to Eq. (\ref{dn}), we can evaluate the cycle difference  $\delta n$ between the sender S and us,  as a function of $X$ and our observational time  $t_d$.
{Geometrically speaking, the numerator of Eq. (\ref{dn}) is the same as the time difference  $[{\rm HR-(CR+CH)}]/c$  (see Fig. 1). For example,  we have  $\sim l_{\rm H}\theta_n/c\sim 1300(\theta_n/{\rm 0.05\,rad})$\,yr for a civilization at the closest approach C.    }

 In  Fig. 5, we present the numerical results, introducing the normalized  distance $x\equiv X/l_{\rm H}$ and fixing $t_d=2024.00$.   Toward the sky direction around the Galactic center, we can keep $\delta n\lsim 5$ to a significant depth $x\lsim 0.5$ along the transmission lines, and the relaxation effects would be relatively unimportant for the signals from these civilizations. {Even specifically, for the $n=3$ sequence,  to keep the observed time difference $P_0\delta n$ less than 80\,yr,  the target civilization should be closer than $0.7\times 8.3=5.8$\,kpc.  
 }

Meanwhile,  by observing the consecutive  pericenter passages, we can directly examine the variation of the orbital periods.   For S2, we will have the next pericenter passage  around $t_0+P_0=t_1=2034.38$ and will make  extensive observation for studying various effects including relativity. { Shortly before the next pericenter passage at $t_1$, we  have the condition $\theta_1\sim 0$ (see Fig. 3) and can  cover the depth $X\sim 1$. The observed period variation will enable us to partly estimate the valid depth $X$  of the target civilizations. If the prospect for the $q=0$ sequence is pessimistic, we could put more weight to the $q=1$ sequence,  which has smaller opening angles $\theta_n$ and is less affected by the long-term orbital perturbation.  }

 Other short-period stars such as S4711 \citep{2022ApJ...933...49P} can also provide us with insights on the {orbital perturbation. Note that the two-body relaxation time $T_{\rm rel}$ will be longer for a smaller semi-major axis   \citep{2010RvMP...82.3121G,2013degn.book.....M}. }
 
 We should also notice that, in  contrast to situation of  a sender, a receiver has  temporal flexibility at data  analysis. For example, in 2036, based on the updated  information of S2's pericenter passage in 2034, we can reanalyze the radio  data taken, e.g. at $t_d=2024.00$.

 \section{Discussion}

 For concreteness,  we have regarded the prominent star S2 as the common reference clock. As stated before, it will be valuable to examine possibilities of other objects, from the view of the Schelling point. 
 Meanwhile,  we might find a peculiar signal  in a blind ETI search around the Galactic center  direction.  Then, we can posteriori   solve  $n$ for  Eq. (\ref{base}), using the timing parameters $(t_0,P_0)$ associated with a small  number  of  prominent reference objects including S2.  {If  the solution $n$ is close to an integer, it may be worth considering to make a deeper search to the incoming direction,  after waiting for  one orbital period $P_0$ of the  corresponding reference object. }
 
  The distribution function for the  durations of technologically  advanced civilizations is totally uncertain. It is also difficult to assess the outlook of ours (see, e.g. \citealt{1993Natur.363..315G}).
 In  the present  scheme with S2,  intentional signals can be repeatedly  transmitted and scanned in a short period $P_0=16$ yr.   This might be advantageous for many Galactic civilizations,  in contrast to  a much longer  scanning interval (e.g.  $\gsim 10^4$ yr  with  binary neutron star mergers as discussed in \citealt{2019ApJ...875L..10S}). 

{
So far, we have not examined relativistic effects to our simplified proposal. In the appendix, we discuss some corrections such as the aberration of light.  While more detailed studies might be required before the actual application of our method, 
we expect that, for  a relatively small $n$ (e.g. $\lsim 10$),  the distance error $\Delta  l_{\rm H}/l_{\rm H}\sim 0.005$ will currently limit our precision (except for the deviation of the clock).  }
  In the future, we might need  to deal with corrections (including those in the appendix) to  improve the total precision, using, e.g. Galactic models.  Some of such effects will share similarities to high-precision measurements including astrometry \citep{2019A&A...625L..10G,2021A&A...647A..59G}.  For the  present scheme, we should also appropriately take into account the Schelling point argument,  if necessary. 
 
 \section{summary}
 The Galactic center is a conspicuous  place in our Galaxy, and,  relatedly, ETI signals have been search around its direction. Many  stars nearly regularly orbit around the massive  black  hole there.   Thanks to the enormous mass of the black hole,   we can determine the distance $l_{\rm H}$ to the Galactic  center at an exceptional  precision ($\Delta l_{\rm H}/l_{\rm H}\sim  0.005$),    despite the remoteness  $l_{\rm H}\sim  8.3$kpc.  
 
 The pericenter passages of a prominent  object such as S2 will be a plausible  option for the reference epoch,  from the view of the Schelling point. Then, with the three locally measured parameters  $(l_{\rm H}, t_0,P_0)$,  individual civilizations can determine the sending and receiving directions and access to the concurrent signaling scheme without prior communication.   Considering the simplicity and usefulness,  this scheme might be at a  Schelling  point in the strategy space  of the interstellar signaling. 
 
 If S2 is really being used as a  reference  clock,   our search directions are  concentric circles  centered by Sgr A$^*$.  The circles  are continuously shrinking  with the period  of $P_0\sim 16$yr (see  Fig. 3).  With our present measurement precision   $\Delta l_{\rm H}/l_{\rm H}\sim  0.005$,  the width  of each circle is $\sim 1'$, and we can compress  the search sky directions  by more than a factor of 100.


\section*{acknowledgments}
The author would like to thank the referee for useful comments to improve the manuscript.


\appendix
 \section{Relativistic Corrections}

\begin{figure}[t]
\vspace{5mm}
 \begin{center}
  \includegraphics[width=10.cm,clip]{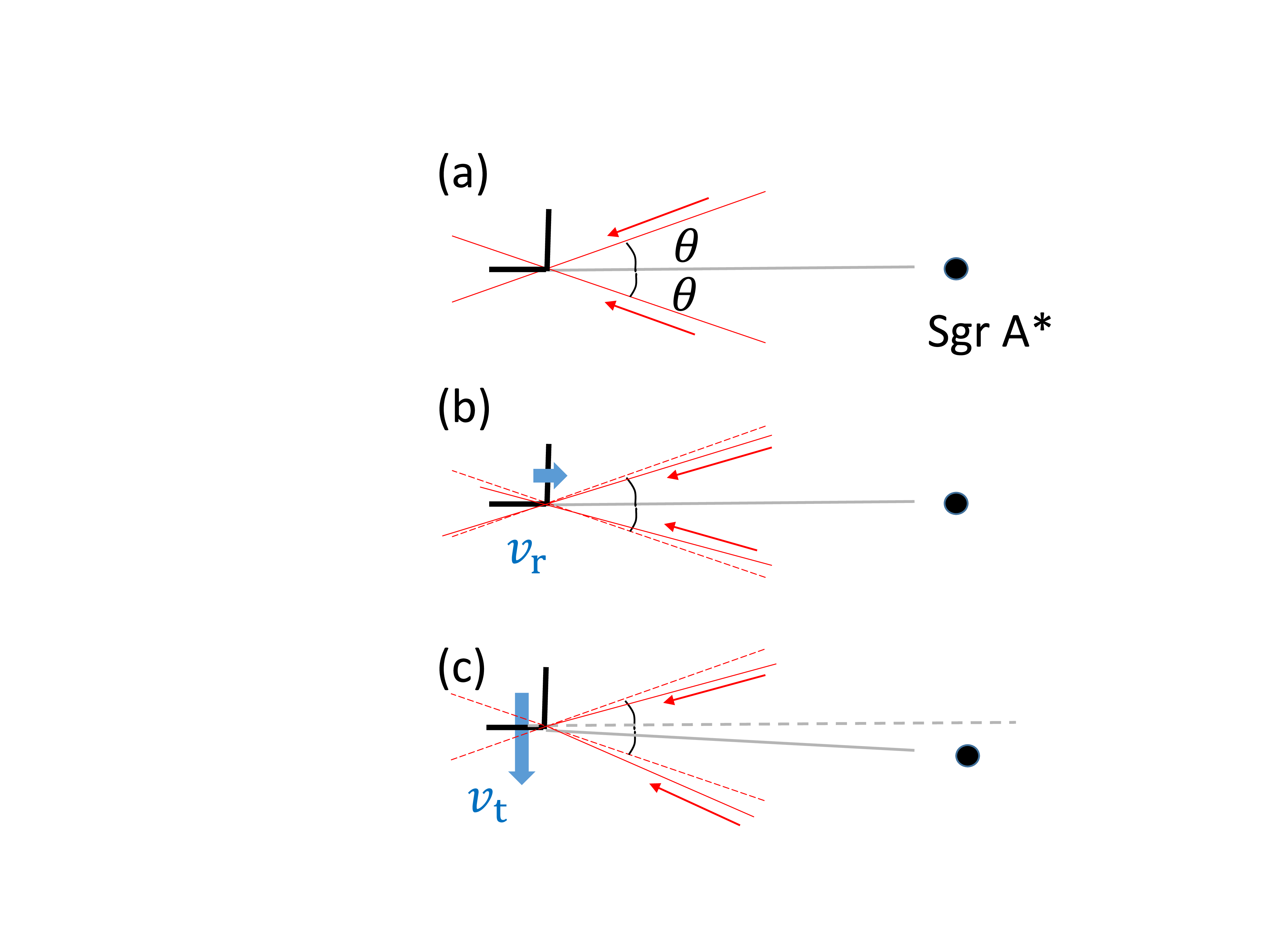}
   \caption{{ Aberration effects on the receiving cones due to the radial and tangential velocities. Frame (a) is at rest frame in the Galaxy.  
   Frame (b) has the radial velocity $v_{\rm r}$ (toward Sgr A*) and the opening angle of its receiving cone (shown with the solid lines) shrinks, compared with the original one (the dashed lines).
   Frame (c) has the transverse velocity $v_{\rm t}$  relative to (a). 
  In frame (c), the apparent sky direction of  the Sgr A* and those of the receiving cone are similarly affected, and the net deviations are evaluated in Eq.  (\ref{net}).  }}
  \label{figure:fig6}
 \end{center}
\end{figure}

 {So far, we have discussed the signaling scheme in a flat space-time, ignoring motions of senders and receivers.  Here we briefly estimate some of the relevant corrections.}
 
{  For our scheme, the simplest  Galaxy-wide frame will be the rest frame  relative to  the Galactic center  (more straightforwardly Sgr A*; see also \citealt{2020ApJ...892...39R}). This frame will be preferable also from the Schelling point.  
To discuss the magnitudes of the  aberration effects caused by the radial and tangential velocities of  the solar system,   we consider the three frames (a), (b) and (c) at the position of the solar system (see Fig. 6). 
Frame (a) is at rest in the Galaxy (excluding the Galactic rotation velocity from the standard frame of rest \citealt{1998gaas.book.....B}).
Relative to  frame (a),  frame (b) has the radial velocity $v_{\rm r}\sim 10{\rm km \, s^{-1}}$ (toward Sgr A*)  and  the frame (c) has the tangential velocity $v_{\rm t} \sim 230{\rm km \, s^{-1}}$ \citep{1998gaas.book.....B} dominated by the Galactic rotation velocity. 
We apply the formula for the aberration angle {(see e.g. Eq. (5.7) in \citealt{1975ctf..book.....L})}
\beq
\delta \phi=\beta_{\rm } \sin \phi \label{abb}
\eeq
 with the angle $\phi$ between the incoming photon and the velocity of a moving frame  ($\beta=v/c$, $v$: the magnitude of the velocity) . This expression is valid for $\delta \phi\ll 1$ and $\beta\ll 1$. }
 
 {
In the second frame (b)  shown in the middle panel of Fig. 6,  the opening angle $\theta(\ll 1)$ of the receiving cone changes by  $\beta_{\rm r}\theta\sim 3\times 10^{-6}(\theta/{\rm 0.1 rad})  $  (setting $\phi=\theta$ and $\beta=\beta_{\rm r}=v_{\rm r}/c$ in Eq. (\ref{abb})).   This correction is much smaller than the uncertainty (\ref{et}) associated with distance error $\Delta l_{\rm H}$.  We should notice that, due to the Doppler effect, in the frame (b), the observed orbital period becomes $(1-\beta_{\rm r})$ times smaller than that in the original frame (a).   With this modified period, for the scheme with $q=0$, the  estimated target angle $\theta$  is shrunk by a factor $\sim(1-\beta_{\rm r}/2)\ne (1-\beta_{\rm r})$ (see, e.g. Eq. (\ref{th0}).  Therefore, unlike the case with $q=1$,   the correction for the radial velocity $v_{\rm r}$ cannot be automatically canceled by using the observed blueshifted orbital period. }

{Next we consider the third frame (c), which has the transverse velocity $v_{\rm t}(=c\beta_{\rm t})$ relative to the frame (a) (see Fig .6). For applying Eq. (\ref{abb}), we evaluate the angles $\phi$ to  the relevant directions.     As shown in the upper panel of Fig. 6,  in the frame (a), Sgr A* is at $\phi=\pi/2$, and,  its receiving cone is in the range   $\phi \in [\pi/2-\theta,\pi/2+\theta]$. 
In the moving frame (c), the three directions $\phi=\pi/2,\pi/2\pm \theta$ are deformed as shown in the solid lines in the bottom panel, following Eq. (\ref{abb}).  Relative to the apparent direction of Sgr A*,  the receiving cone has the maximum deformation
\beq
|\delta \phi_{\pi/2\pm \theta}-\delta \phi_{\pi/2}|\sim \frac{\beta_{\rm t} \theta^2}2\sim 10^{-5}\lmk  \frac{\theta}{\rm 0.1 rad} \rmk^2,  \label{net}
\eeq
which is also smaller than the uncertainty (\ref{et})  originating from the distance error $\Delta l_{\rm H}$.  }

{In our study, we studied propagation of photons in a flat space-time, without considering the Galactic potential. Associated with the path length difference in the context of Sec. 5 (e.g.  $\delta l={\rm RS+SH-HR}$  in Fig. 1), we actually have a relativistic time correction (Shappiro time delay). We can roughly estimate its magnitude by 
\beqa
\frac{\beta^2\delta l}c&\sim& 1\lmk  \frac{\delta l}{\rm 1kpc}\rmk {\rm day}\\
& & \nonumber
\eeqa
 with the factor $\beta$ for the Galactic rotation velocity $200-300{\rm km\,s^{-1}}$ (see also Fig. 1 in \citealt{2016MPLA...3150083D}).  This correction will be smaller than the uncertainty $\sim 30$ days, corresponding to the distance error $\Delta l_{\rm H}$ (see the paragraph after Eq. (\ref{et})).
}

\end{document}